\documentclass[aps,prl,twocolumn,showpacs,superscriptaddress]{revtex4-1}
\usepackage{mathtools,amssymb,lipsum, nccmath}
\usepackage{bm}

\usepackage{graphicx} 
\usepackage{color}
\usepackage{float}

\usepackage{booktabs} 
\usepackage{array} 
\usepackage{paralist} 
\usepackage{verbatim} 
\usepackage{subfig} 
\usepackage{float} 
\usepackage{hyperref}
\hypersetup{colorlinks=true,linkcolor=blue,urlcolor=blue}

\usepackage{hyperref}

\usepackage{listings}
\usepackage{xcolor}
\usepackage{listings}
\usepackage{pdfpages} 
\usepackage{pgffor} 
\usepackage{xr} 

\makeatletter
\AtBeginDocument{\let\LS@rot\@undefined}
\makeatother

\def\supplementfilename{supplement}

\pdfximage{\supplementfilename.pdf}
\def\numbersupplementpages{\the\pdflastximagepages}

\newif\ifarXiv
\arXivtrue 

\begin{document}

\title{Floquet driven frictional effects}

\author{Vahid Mosallanejad}
\email{vahid@westlake.edu.cn} 
\affiliation{Department of Chemistry, School of Science, Westlake University, Hangzhou, Zhejiang 310024, China}
\affiliation{Institute of Natural Sciences, Westlake Institute for Advanced Study, Hangzhou, Zhejiang 310024, China}

\author{Jingqi Chen}
\affiliation{Department of Chemistry, School of Science, Westlake University, Hangzhou, Zhejiang 310024, China}
\affiliation{Institute of Natural Sciences, Westlake Institute for Advanced Study, Hangzhou, Zhejiang 310024, China}

\author{Wenjie Dou}
\email{douwenjie@westlake.edu.cn} 
\affiliation{Department of Chemistry, School of Science, Westlake University, Hangzhou, Zhejiang 310024, China}
\affiliation{Department of Physics, School of Science, Westlake University, Hangzhou, Zhejiang 310024, China}
\affiliation{Institute of Natural Sciences, Westlake Institute for Advanced Study, Hangzhou, Zhejiang 310024, China}

\begin{abstract}
When the coupled electron-nuclear dynamics are subjected to strong Floquet driving, there is a strong breakdown of the Born-Oppenheimer approximation. In this article, we derive a Fokker-Planck equation to describe non-adiabatic molecular dynamics with electronic friction for Floquet driven systems. We first provide a new derivation of the Floquet quantum-classical Liouville equation (QCLE) for driven electron-nuclear dynamics. We then transform the Floquet QCLE into a Fokker-Planck equation with explicit forms of frictional force and random force. We recast the electronic friction in terms of Floquet Green's functions such that we can evaluate the electronic friction explicitly. We show that the Floquet electronic friction tensor exhibits antisymmetric terms  even at equilibrium for real-valued Hamiltonian, suggesting that there is a Lorentz-like force in Floquet driven non-Born Oppenheimer dynamics even without any spin-orbit couplings. 
\end{abstract}
\maketitle

\textit{Introduction.--}
The molecular dynamics near metallic surfaces can be non-adiabatic in nature and hence Born-Oppenheimer (BO) approximation is not necessary correct \cite{rittmeyer2018energy,galperin2015nuclear,juaristi2008role}. The electronic friction approach is considered as the first order correction to the BO approximation \cite{dou2018perspective}, which can be understood as quantum mechanical damping force of a manifold of fast relaxing electronic on classical nuclear motion. Electronic friction approaches were successful in explaining many experimental results such as molecular beam experiments \cite{spiering2018testing, lu2019semi, hertl2021random}, electrochemistry \cite{lam2019theory}, charge$/$spin transport phenomena\cite{dou2018broadened,bajpai2020spintronics}.  Quantitatively, electronic friction is a tensor which appears on the generalized Langevin equation \cite{askerka2016role}. One of the first notable quantum mechanical derivations of the electronic friction tensor is given by Head-Gordon and Tully \cite{head1995molecular}. Later, more rigorous expressions are derived from Keldysh Green's function \cite{bode2012current,kershaw2020non}, path integral \cite{lu2012current,chen2019electronic}, quantum classical Liouville equation (QCLE) \cite{dou2016many}, exact factorization \cite{martinazzo2022quantum}. It has been shown that there is only one universal electronic friction tensor in the Markovian limit \cite{dou2017universality,dou2018universality}. Furthermore, study shows that the friction tensor can exhibit antisymmetric terms even at equilibrium when spin-orbit couplings are involved. \cite{teh2021antisymmetric}  

Now, there are increasing interests in understanding of the dynamics of molecular systems with strong light-matter interactions, which is helpful for interpreting photochemistry and spectroscopy \cite{schiro2021quantum,lang2015dynamical}. In particular, people are interested in how to use light$/$photon to manipulate chemical reactions where the dynamical interplay between light and electronic non-adiabatic transitions plays a significant role. At the same time, active research is currently ongoing to understand the response of quantum systems to a periodic driving force, or so called ``Floquet driven" systems \cite{sambe1973steady, kohler2005driven, novivcenko2017floquet}. Floquet theorem provides a powerful method for the analysis of quantum systems subjected to periodic external drivings. Effects, such as phase transitions and pump-probe photoemission can be explained by applying Floquet theorem in solving quantum mechanical problems \cite{yang2019floquet, dehghani2014dissipative,sentef2015theory}. The coupled electron-nuclear dynamics with strong light-matter interactions  can be described by the Floquet quantum classical Liouville equation (QCLE) successfully. \cite{kapral2006progress,kelly2010quantum,kim2008quantum}. In this article, we offer a new derivation for the Floquet QCLE starting from Floquet Liouville equation. Moreover, we map the Floquet QCLE into a Langevin equation with all non-adiabatic correction being incorporated into frictional effects. Furthermore, we demonstrate that the Floquet electronic friction tensor exhibits antisymmetric terms  even at equilibrium for real-valued Hamiltonian. 

\textit{Liouville-von Neumann equation in the Floquet representation.--} 
For the coupled electron-nuclear motion, we consider a general Hamiltonian $\hat H$ that can be divided into the electronic Hamiltonian $\hat H_e$ and the nuclear kinetic energy: 
\begin{equation}
\label{e1}
\hat{H}  =\hat{H}_e ({\bold R}, t) +\sum_{\alpha} \frac{\hat{P}_{\alpha}^{2}}{2 M_{\alpha}}  
\end{equation}
Here ${\mathbf{R}}=\{{R}_{\alpha}\}$ and $\hat{\mathbf{P}}=\{\hat{P}_{\alpha}\}$ are position and momentum operators for the nuclei respectively. We use $\alpha$ to denote nuclear degrees of freedom. Note that, the electronic Hamiltonian $\hat{H}_e ({\bold R}, t)$ is considered to be an explicit function of ${\mathbf{R}}$ and time $t$. Below, we will consider the case that the system is subjected to periodic driving, such that $\hat{H}_e ({\bold R}, t+T) = \hat{H}_e ({\bold R}, t)$. $T$ is the period of the driving frequency. 

The equation of motion for the density operator follows Liouville-von Neumann (LvN): $\frac{d\hat{\rho}(t)}{dt} = -\frac{i}{\hbar} [\hat{H} (t), \hat{\rho}(t)]$. For the periodic driving system, we can derive a Floquet Liouville-von Neumann (LvN) equation describe the time evolution of the density operator in Floquet representation. To do so, two transformations are needed to derive Floquet representation of LvN: (I) Transformation of LvN into the Fourier representation and (II) transformation from the Fourier representation to the Floquet representation. The details of part (I) is given in the Supplementary Information S1. Fourier representation of the LvN equation reads
\begin{eqnarray} 
\begin{aligned}
\label{e2}
&\frac{d\hat{\rho}^{f}(t)}{dt} 
= -\frac{i}{\hbar} [\hat{H}^{f}(t), \hat{\rho}^{f}(t)],
\end{aligned}
\end{eqnarray}
Here, the Fourier representations of Hamiltonian and density operators [$\hat{H}^{f}(t)$ and $\hat{\rho}^{f}(t)$] are  given by
\begin{eqnarray}
\label{e3}
\hat{H}^{f}(t)=\sum_{n}^{} \hat{H}^{(n)} \hat{L}_{n}  e^{i n \omega t}, 
\hat{\rho}^{f}(t)=\sum_{n}^{} \hat{\rho}^{(n)} (t) \hat{L}_{n} e^{in\omega t}.~~
\end{eqnarray} 
The operator $\hat{L}_{n}$ denotes the $n$th ladder operator in Fourier space (see SI for detailed definition). The Fourier expansion coefficients of the Hamiltonian is given by $\hat{H}^{(n)}=1/T\int_0^TH(t)e^{-in\omega t}dt$. Indeed, Eqs. (3) are Fourier expansions modified by adding the ladder operator $\hat{L}_{n}$. We stress that, the ladder operator turns the vector-like Fourier expansion into a matrix-like representation. 
We then transform the density operator from the Fourier representation to the Floquet representation as
\begin{eqnarray} 
\label{e4}
\hat{\rho}_{F}(t)=e^{-i \hat{N} \omega t} \hat{\rho}^{f} (t) e^{i \hat{N} \omega t}=
\sum_{n}^{} \hat{\rho}^{(n)} (t) \hat{L}_{n}, 
\end{eqnarray}
where $\hat{N}$ is the number operator in Floquet representation (see SI for detailed definition). Employing such a definition, the equation of motion for $\hat{\rho}_{F}(t)$ now reads 
\begin{eqnarray}
\label{e5}
\frac{d}{d t} \hat{\rho}_{F}(t)= -\frac{i}{\hbar} \left[\hat{H}_{F}, \hat{\rho}_{F}(t)\right],
\end{eqnarray}
where we have defined the following Floquet representation for the Hamiltonian as
\begin{eqnarray} 
\label{e6}
\begin{aligned}
\hat{H}_{F}
&= \sum_{n}^{} \hat{H}^{(n)} \hat{L}_{n}+\hat{N} \hbar \omega.
\end{aligned}
\end{eqnarray}
We have used the commutation relations between the ladder and number operators, $[\hat{N}, \hat{L}_{n} ] =n \hat{L}_{n}$ and $e^{-i \hat{N} \omega t} \hat{L}_{n} e^{i \hat{N} \omega t}= \hat{L}_{n}e^{-i n \omega t}$ to derive the above equations. We note that the Floquet LvN equation have the same structure as the traditional LvN. The advantage of the Floquet LvN is to allow us to program the dynamics using the time independent Hamiltonian.  
\textit{Floquet QCLE.--}
To derive the Floquet QCLE, we perform the partial Wigner transformation with respect to the nuclear degrees of freedom on the Floquet LvN equation, Eq. (5), as
\begin{eqnarray}
\label{e7}
\frac{d}{d t} (\hat{\rho}_{F})_W 
(\bold R,\bold P, t)
= -\frac{i}{\hbar} \left(
(\hat{H}_{F}\hat{\rho}_{F})_W
-
(\hat{\rho}_{F}\hat{H}_{F})_W
\right).~
\end{eqnarray}
We have used subscript $W$ to denote the Wigner transformation. The Wigner transformation is given by 
\begin{eqnarray}
\begin{aligned}[t]
\label{e8}
\hat{O}_{\rm W} \left( \bm{R}, \bm{P}, t \right)
\equiv
\int d\bm{Y} 
e^{\frac{-i\bm{R}\cdot \bm{P}} {\hbar}}
\langle \bm{R} - \frac{\bm{Y}}{2} |\hat{O}(t) | \bm{R} + \frac{\bm{Y}}{2} \rangle,~~~~
\end{aligned}
\end{eqnarray}
where $\hat{O}(t)$ is an arbitrary operator and $|\bold R \rangle $ is the real space representation of the nuclear degree of freedom. As a result of this transformation, $\bold R$ and $\bold P$ can be interpreted as position and momentum variables in the classical limit. Note that, the Wigner-Moyal operator can be used to express the partial Wigner transform of the product of operator $\hat A$ and $\hat B$: 
\begin{eqnarray}
\label{e9}
\begin{aligned}
&(\hat{A} \hat{B})_{W}(\bold R,\bold P)=
\hat{A}_{W}(\bold R,\bold P)   e^{-i \hbar \overleftrightarrow{\Lambda} / 2}    \hat{B}_{W}(\bold R,\bold P), \\
&\overleftrightarrow{\Lambda}=
\sum_{\alpha} 
\overleftarrow{ \frac{\partial}{ \partial P^{\alpha}} }    \overrightarrow{  \frac{\partial}{\partial R^{\alpha}} }-
\overleftarrow{ \frac{\partial}{\partial R^{\alpha}}  }    \overrightarrow{ \frac{\partial}{\partial P^{\alpha}} }.
\end{aligned}
\end{eqnarray}
When truncating the Wigner-Moyal operator to the first order in the Tayler expansion, $e^{-i \hbar \overleftrightarrow{\Lambda} / 2}\approx (1-i \hbar \overleftrightarrow{\Lambda} / 2)$, we arrive at the Floquet QCLE as 
\begin{eqnarray}
\label{e10}
\begin{aligned}
&\frac{d}{d t} \hat{\rho}_{WF} (\bold R,\bold P, t)=
- i/\hbar\left[\hat{H}_{WF}, \hat{\rho}_{WF}(t)\right]\\
&-\frac{1}{2}
\left(   
 \hat{H}_{WF} \overleftrightarrow{\Lambda}  \hat{\rho}_{WF}
 -\hat{\rho}_{WF}  \overleftrightarrow{\Lambda}   \hat{H}_{WF} 
\right),
\end{aligned}
\end{eqnarray}
Here, we have denoted $(\hat{O}_{F})_W(\bold R,\bold P) \equiv \hat{O}_{WF} ({\bold R},{\bold P})$. The subscript $\mathop{WF}$ indicates that the Wigner transformation performed after the Floquet transformation.  
For the coupled electron-nuclear Hamiltonian in Eq. (1), we can rewrite the Floquet QCLE as follows
\begin{eqnarray}
\label{e11}
\begin{aligned}
&\frac{\partial}{\partial t} \hat{\rho}_{WF}(t)=
-\hat{\hat{\mathcal{L}}}_{WF}(\hat{\rho}_{WF}(t))
-\sum_{\alpha}    \frac{P_{\alpha}}{M_{\alpha}}  \frac{\partial \hat{\rho}_{WF}(t)}{\partial R_{\alpha}}\\
&+\frac{1}{2}     \sum_{\alpha}   \left\{\frac{\partial \hat{H}^{e}_{WF}}{\partial R_{\alpha}}, \frac{\partial \hat{\rho}_{WF}(t)}{\partial P_{\alpha}} \right\}.
\end{aligned}
\end{eqnarray}
Here $\hat{\hat{\mathcal{L}}}_{WF}(\hat{\rho}_{WF}(t))\equiv i/\hbar[\hat{H}^e_{WF},\hat{\rho}_{WF}(t)]$. $\hat{H}^{e}_{WF}$ is the Floquet-Wigner transformed electronic Hamiltonian $H_e$. We have also denoted the anti-commutator as $\{\hat{A},\hat{B}\}=\hat{A}\hat{B}+\hat{A}\hat{B}$. This Floquet QCLE is consistent with the recently published work (see Eq. 14 in Ref. \cite{chen2020proper}). Such a Floquet QCLE represents the non-adiabatic dynamics of the coupled electron-nuclear motion subjected to periodic driving. 
\textit{The Fokker-Planck equation.--} In the limit when the nuclear motion is slow as compared to electronic motion as well as the driving speed, we can trace out all electronic degrees of freedom and Floquet levels, such that we are left with the pure nuclear density. To be more explicit, $\mathcal{A}(\bold R,\bold P,t) = Tr_{e, F}  \hat{\rho}_{WF} $.  Here, $Tr_{e,F}$ denotes trace over both many-body electronic states and Fourier space. To the first order in the correction to the BO approximation, we arrive at a Fokker-Planck equation for the pure nuclear density $A$: 
\begin{eqnarray}
\label{e12}
\begin{aligned}
&\frac{\partial }{\partial  t}\mathcal{A}=
-\sum_{\alpha} 
\frac{P_{\alpha}}{m_{\alpha}} 
\frac{\partial \mathcal{A} }{\partial  R_\alpha}
-\sum_{\alpha} F_{\alpha} \frac{\partial \mathcal{A}}{\partial P_{\alpha}} +\\
&\sum_{\alpha \beta} \gamma_{\alpha \beta} \frac{\partial}{\partial P_{\alpha}}
\left(\frac{P_{\beta}}{m_{\beta}} \mathcal{A}\right)
+\sum_{\alpha \beta} \bar{D}_{\alpha \beta}^{S} \frac{\partial^{2} \mathcal{A}}{\partial P_{\alpha} \partial P_{\beta}}.
\end{aligned}
\end{eqnarray}
The detailed derivation can be found in the SM. Here $F_{\alpha} = Tr_{e, F} ({\partial \hat{H}_{F}^{e}}/{\partial R_{\alpha}}  \hat{\rho}_{s s F} )$ is the mean force. $\hat{\rho}_{s s F}$ is the steady state Floquet electronic density. The Fokker-Planck equation is equivalent to the Langevin equation 
\begin{equation} 
\label{e13}
m_{\alpha} \ddot {R}_\alpha=F_\alpha-\sum_\beta \gamma_{\alpha\beta} \dot R_\beta +  \delta F_\alpha.
\end{equation}
Here $\delta F_\alpha $ is the random force, which satisfies $ \frac12 \langle \delta F_\alpha (0) \delta F_\beta  (t) + \delta F_\beta (0) \delta F_\alpha (t) \rangle = \bar{D}_{\alpha \beta}^{S}\delta (t) $. $\bar{D}_{\alpha \beta}^{S}$ is the correlation function of the random force. $\gamma_{\alpha \beta}$ is the friction coefficient 
\begin{eqnarray}
\label{e14}
\gamma_{\alpha \beta}=
-\int_{0}^{\infty} dt Tr_{e, F}
\left(
\frac{\partial \hat{H}_{F}^{e}}{\partial R_{\alpha}} e^{\frac{-i \hat{H}_{F}^{e} t }{ \hbar}}
 \frac{\partial \hat{\rho}_{ssF}}{\partial R_{\beta}} e^{\frac{i \hat{H}_{F}^{e} t }{ \hbar}}
\right).~~ 
\end{eqnarray}
Up to now, we have successfully transformed the coupled electron-nuclear motion subjected to periodic driving into a Langevin equation for the pure nuclear motion with all electronic motion and Floquet driving being incorporated into frictional force and random force. We now proceed to evaluate the frictional force in terms of Green's function. 

\textit{Quadratic electronic Hamiltonian.--} 
Our derivation above is general as long as the nuclear motion is slow as compared to electronic motion and Floquet driving. We now consider quadratic electronic Hamiltonian, 
\begin{eqnarray}
\label{e15}
\hat{H}_e ({\bold R}, t)  = \sum_{ab} \mathcal{H}_{ab} ({\bold R},  t) \hat b^\dagger_a \hat b_b.
\end{eqnarray}
One can then transform the Floquet electronic friction into the single particle representation as 
\begin{eqnarray}
\label{e16}
\begin{aligned}
\gamma_{\alpha \beta}=
- \hbar \int_{-\infty}^{\infty} 
\frac{d \epsilon}{2 \pi} ~Tr_{m,F}
\left(
\frac{\partial \mathcal{H}_{F}}{\partial R_{\alpha}}  
G_F^{R}
\frac{\partial \sigma_{ssF} }{\partial R_{\beta}} 
G_F^{A}
\right),
\end{aligned}
\end{eqnarray}
where $Tr_{m,F}$ denotes the trace over both single particle electronic DoFs and Fourier space. We have defined the Floquet Retarded and Advance Green's function: $G_F^{R/A}=(\epsilon\pm i\eta-\mathcal{H}_{F})^{-1}$, $\eta\rightarrow 0^+$. Here, $\mathcal{H}_{F}$ is the Floquet single particle electronic Hamiltonian, and ${\sigma}_{{ssF}}$ denotes the Floquet single-particle density matrix, which is defined as $[{\sigma}_{ssF}]_{ab} =Tr_{e}(\hat{b}_{b}^{\dagger} \hat{b}_{a} \hat{\rho}_{ssF})$. The Floquet single-particle density matrix can be further expressed in terms of Floquet lesser Green's function, such that the final expression for the Floquet electronic friction is given by:  
\begin{eqnarray}
\label{e17}
\begin{aligned}
\gamma_{\alpha \beta}=
\hbar \int_{-\infty}^{\infty} 
\frac{d \epsilon}{2 \pi} ~ Tr_{m,F}
\left(
\frac{\partial \mathcal{H}_{F}}{\partial R_{\alpha}}  
\frac{\partial G_F^{R} }{\partial \epsilon} 
\frac{\partial \mathcal{H}_{F}}{\partial R_{\beta}}  G_F^{<} \right)+ \textrm{h.c}. ~~~~
\end{aligned}
\end{eqnarray}
See the Supplementary Information for the details of derivation. $G_F^{<}$ is the lesser Floquet Green's function. Note that the Floquet electronic friction is the same as non-Floquet electronic friction, except Green's functions are now the Floquet version of the corresponding Green's function. 
\textit{Dot-lead separation.--}
We now consider a specific model, such that we can calculate the Floquet electronic friction explicitly. We will demonstrate that the Floquet driving electronic friction exhibits anti-symmetric terms for real Hamiltonian even without any current. To be more specific, we consider a Hamiltonian with dot-lead separation:    
\begin{eqnarray}
\label{e18_21}
\hat{H}_e &=& \hat H_s  + \hat H_b + \hat H_v   \\
\hat H_s  &=& \sum_{ij} [h^s]_{ij} (\bold R, t) \hat d_i^\dagger \hat d_j + U(\bold R) \\
\hat H_b &=& \sum_{\zeta k} \epsilon_{\zeta  k} \hat c_{\zeta k}^\dagger \hat c_{\zeta k}  \\
\hat H_v  &=&  \sum_{\zeta k,i} V_{\zeta k,i}  (   \hat c_{\zeta k}^\dagger \hat d_i + \hat d^\dagger_i  \hat c_{\zeta k} ) 
\end{eqnarray}
Here, $\hat H_s$ is the dot Hamiltonian. The bath Hamiltonian consists of the left and right ($\zeta = L, R$) leads. $\hat H_v$ describes the system-bath couplings. $U(\bold R)$ is the potential for the nuclei. 

For such a model, we can calculate Floquet Green's function exactly. In particular, the Retarded Green's function for the system is given by:
\begin{equation}
\label{e22}
G_{sF}^{R}(\epsilon) =
\left( 
\epsilon  - \Sigma^R_F(\epsilon) -h^s_{F}
\right)^{-1},
\end{equation}
$\Sigma^R_F(\epsilon) = \sum_{\zeta = L, R} \Sigma_{\zeta F}^{R}$ is the total self energy in Floquet representation. The elements of the self energy is given by $[\Sigma_{\zeta F}^{R}]_{ij}(\epsilon) =  \sum_{k} V_{\zeta k, i} g_{F, \zeta k}^R(\epsilon) V_{\zeta k, j}$, where the $g_{F, \zeta k}^R(\epsilon)=(\epsilon+i 0^+ -\epsilon_{\zeta k}- \hat{N}  \hbar \omega)^{-1}$ is the $k$-th element of the Retarded Green's function of the isolated lead $\zeta$. $h^s_{F}$ is the Floquet representation of the dot energy level.  The lesser Green's function for the system is then given by 
\begin{eqnarray} 
\label{e23}
G_{sF}^{<}(\epsilon)=G_{sF}^{R} (\epsilon) \Sigma_{F}^{<} (\epsilon)  G_{sF}^{A} (\epsilon),
\end{eqnarray}
Here, $\Sigma_{F}^{<}(\epsilon)= \sum_{\zeta = L, R} \Sigma_{\zeta F}^{<}(\epsilon)$ is the lesser Green's function, which can be evaluated as $[\Sigma_{\zeta F}^{<}]_{ij}(\epsilon)\equiv \sum_{k} V_{{\zeta k},i} g_{F, {\zeta k}}^<(\epsilon) V_{{\zeta k},j}$. Here, $g_{F, {\zeta k}}^< (\epsilon)$ is the lesser green's function for the $\zeta$ lead.  $g_{F, {\zeta k}}^< (\epsilon) =i2\pi f(\epsilon- \hat{N}  \hbar \omega- \mu_\zeta)\delta(\epsilon - \epsilon_{\zeta k}- \hat{N}  \hbar \omega)$ where $f$ is the Fermi function. In what following, we will invoke the wide band approximation, such that $[\Sigma_{\zeta F}^{R}]_{ij}(\epsilon) = -\frac i 2 \Gamma_{ij} $, and $[\Sigma_{\zeta F}^{<}]_{ij}(\epsilon) =  i  \Gamma_{ij} f(\epsilon - \hat N \hbar\omega - \mu_\zeta ) $. We can then proceed to calculate Floquet electronic friction using these Green's functions. 

\textit{Results and Discussions.--} 
We will now consider a two-level and two nuclear DoFs model:
\begin{eqnarray}
\label{e24}
[h^s] (x,y,t)= 
\left(
\begin{array}{cc}
x+\Delta & Ay+Bcos(\omega t) \\
A y+B cos(\omega t) & -x-\Delta 
\end{array} 
\right).~~
\end{eqnarray}
The nuclear potential $U(\bold R)$ is taken to be harmonic oscillators in both $x$ and $y$ dimensions. The diagonal terms of Hamiltonian represent two shifted parabolas in $x$ direction with a driving force of $2\Delta$. The off-diagonal couplings depend on displacement in $y$ direction as well as external time-periodic driving $Bcos(\omega t)$ from a monochromatic light source. $B$ represents the strength of the external driving (e.g., the intensity of light) and $\omega$ is the frequency of the time-periodic driving. Below, we consider the case where the first level couples to the left lead and the second level couples to the right lead, and we set $\Gamma_{11} = \Gamma_{22} = \Gamma$.   
In the equilibrium case (where $\mu_L = \mu_R$) and without any driving, the electronic friction is shown to be symmetric along nuclear DoFs provided the Hamiltonian is real \cite{teh2021antisymmetric}. In Fig. 1, we plot the friction tensor as a function of the nuclear coordinates $(x,y)$. In particular, we define the symmetric and antisymmetric components [$\gamma^S_{xy}=(\gamma_{xy}+\gamma_{yx})/2, \gamma^A_{xy}=(\gamma_{xy}-\gamma_{yx})/2$] of the friction tensor. In the absence of external driving ($B=0$), the antisymmetric component is indeed vanished (as predicted). The frictions tensors $\gamma_{xx}$ and $\gamma_{yy}$ consists of two Gaussian curves which are merged along the orientations of the nuclear coordinate. This results agree with previous findings for real Hamiltonian without any driving \cite{teh2021antisymmetric}.  
\begin{figure}[h!tb]
  \centering
  \includegraphics[width =7.5cm,height=7.5cm]{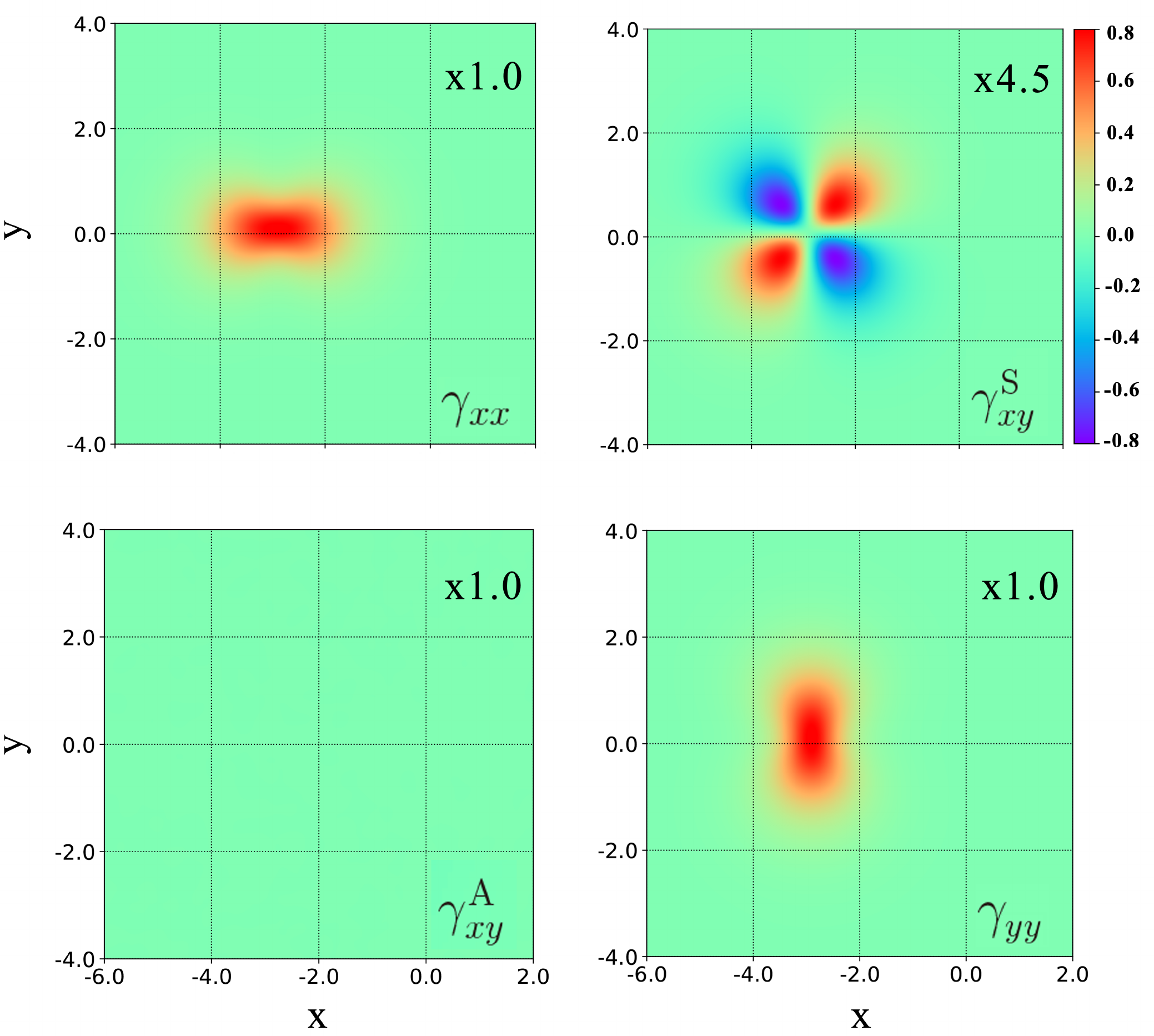}
  \caption{\label{fig1} Floquet friction tensors in absence of external driving $B = 0$: $\gamma_{xx}$ (top left), $\gamma_{xy}^S$ (top right), $\gamma_{xy}^A$ (bottom left) and $\gamma_{yy}$ (bottom right). Parameters: $\Gamma $=1, $\mu_{R,L}=0$, $\beta= 2$, $A = 1$, $\Delta= 3$, $\omega= 0.5$. We have used $N=5$ Floquet levels to converge the results.} 
  \end{figure}
\begin{figure}[h!tb]
  \centering
  \includegraphics[width=7.5cm,height=7.5cm]{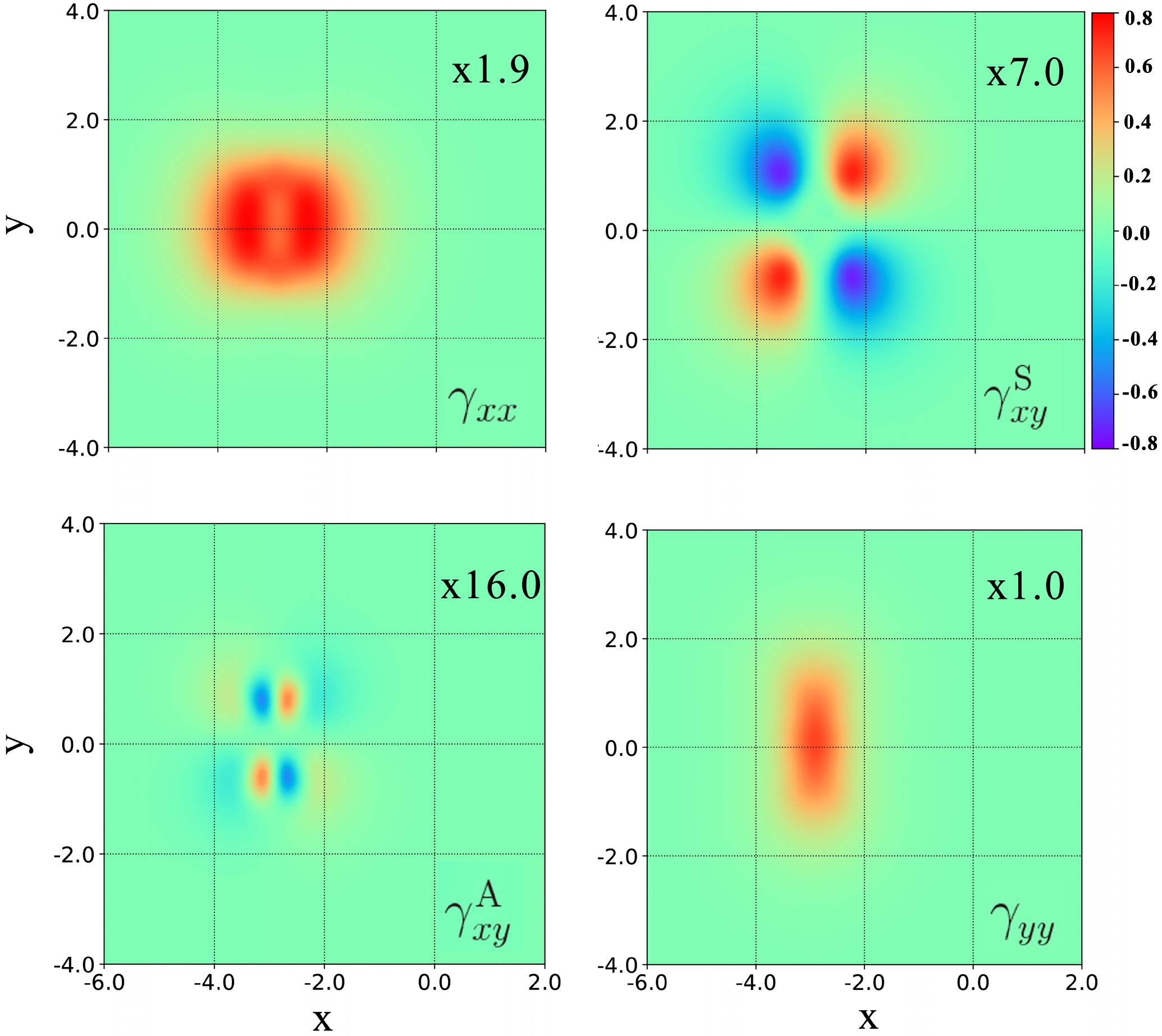}
  \caption{\label{fig2} Floquet friction tensors in presence of external driving:  $\gamma_{xx}$ (top left), $\gamma_{xy}^S$ (top right), $\gamma_{xy}^A$ (bottom left) and $\gamma_{yy}$ (bottom right). Parameters: $\Gamma $=1, $\mu_{R,L}=0$, $\beta= 2$, $A = 1$, $\Delta= 3$, $\omega= 0.5$, $B=1$, $N=5$.} 
 \end{figure}
We  now turn on  time-periodic off-diagonal coupling by setting $B=1$. As shown in Fig. 2, the antisymmetric term $\gamma_{xy}^A$ is no longer zero when Floquet driving is turning on. Moreover, the distributions of $\gamma_{xx}$, $\gamma_{yy}$, and $\gamma_{xy}^S$  in the real space is enlarged  as compared to the non-Floquet case. The magnitude of $\gamma_{xx}$ and $\gamma_{xy}^S$ are also increased by almost factor of 2. 
Finally, in Fig. 3 we plot the frictional terms for the increased driving frequency ($\omega=1$). In such a case, the magnitude of the antisymmetric terms (the Lorentz force) is notably increased, whereas the magnitudes of the other terms do not change significantly. Interestingly, the shape of $\gamma_{xx}$ is composed of two large ellipses and two small ones. The central distance between the larger ellipse and the smaller one  in $x$ axis is about $\omega$. This is consistent with the picture of Floquet replica of the potential surfaces separated by $\omega$. Finally, note that all friction terms have mirror symmetry around the avoided crossing point ($x = -\Delta$ and $y = 0$) and magnitudes of $\gamma_{xy}^A$ and $\gamma_{xy}^S$ are always maximized far from the avoided crossing. 
\begin{figure}[h!tb]
  \centering
  \includegraphics[width=7.5cm,height=7.5cm]{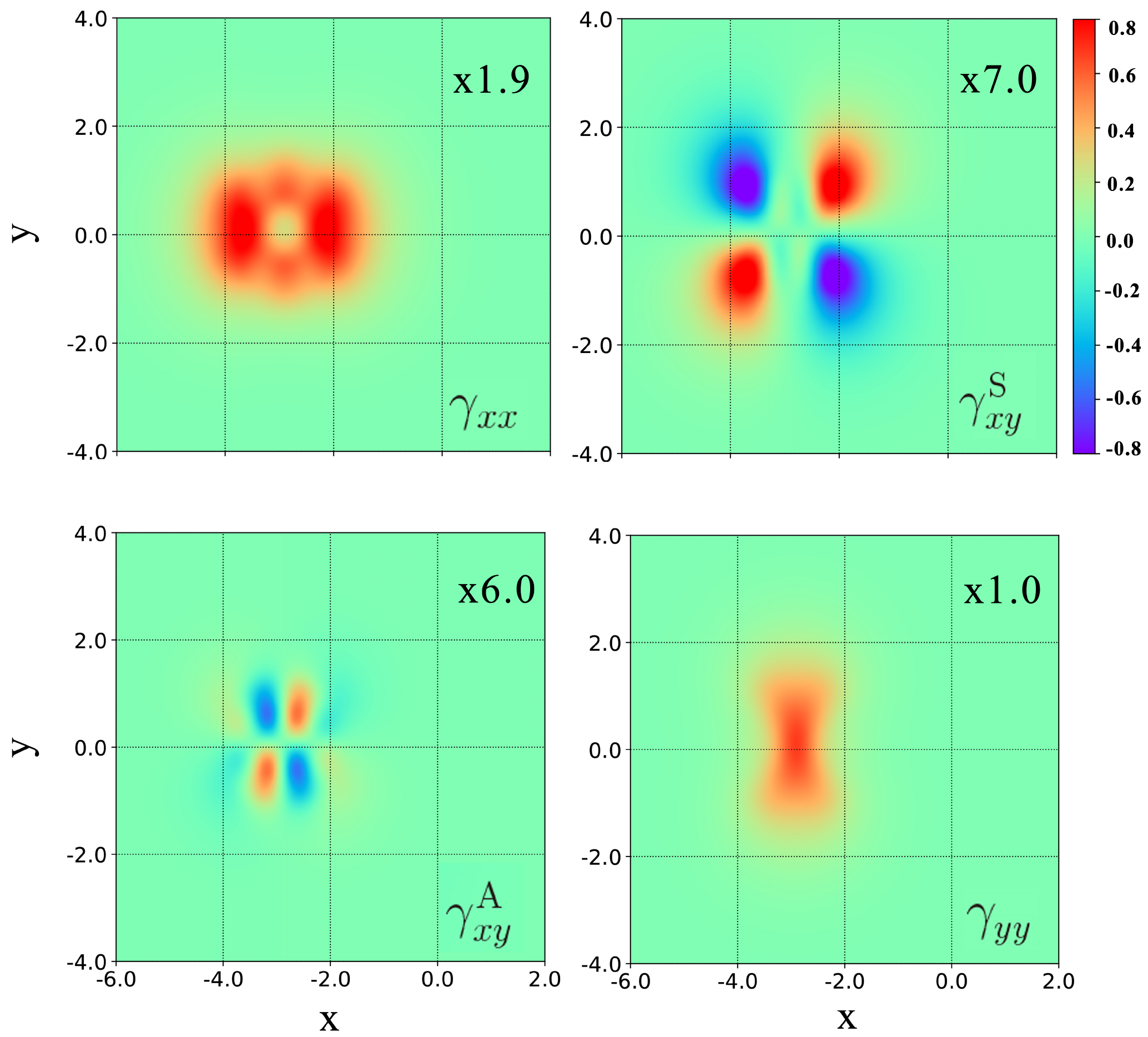}
  \caption{\label{fig3} Floquet Friction tensors in presence of external driving with a larger frequency:  $\gamma_{xx}$ (top left), $\gamma_{xy}^S$ (top right), $\gamma_{xy}^A$ (bottom left) and $\gamma_{yy}$ (bottom right). Parameters: $\Gamma $=1, $\mu_{R,L}=0$, $\beta= 2$, $A = 1$, $\Delta= 3$, $\omega= 1$, $B=1$, $N=5$.} 
  \end{figure}
\textit{Conclusion.--} 
We have formulated quantum--classical Liouville equation in Floquet representation to describe non-adiabatic dynamics with light-matter interactions. We have further mapped the Floquet QCLE into a Langevin dynamics where all electronic DoFs and light-matter interactions are incorporated into a friction tensor. We then recast the friction tensor into the form of Floquet Green's functions such that we can evaluate the friction tensor explicitly. We show that the light-matter interactions can introduce anti-symmetric friction tensor even at equilibrium without any spin-orbit couplings. Future work must explore how the Lorentz-like  force affects the dynamics in a realistic situation. 
We acknowledge the startup funding from Westlake University.  
\bibliography{FleElecFricBIB}


\ifarXiv
    

\section*{Supplementary material}

# Supplementary Information for Floquet driven frictional effects

Vahid Mosallanejad,[1,2,*] Jingqi Chen,[1,2] and Wenjie Dou[1,3,2,†]

[1]*Department of Chemistry, School of Science, Westlake University, Hangzhou, Zhejiang 310024, China*
[2]*Institute of Natural Sciences, Westlake Institute for Advanced Study, Hangzhou, Zhejiang 310024, China*
[3]*Department of Physics, School of Science, Westlake University, Hangzhou, Zhejiang 310024, China*

PACS numbers:

## DRIVATION OF LIOUVILLE-VON NEUMANN EQUATION IN THE FOURIER REPRESENTATION

In what follows we show that Liouville-von Neumann (LvN) equation of motion in the Fourier representation keeps the same form as the non-Fourier one as: $d\hat{\rho}^f(t)/dt = -i/\hbar[\hat{H}^f(t), \hat{\rho}^f(t)]$ where $\hat{\rho}^f$ and $\hat{H}^f$ denote the density and Hamiltonian operators in the Fourier representation [S1]. The procedure has two parts; 1) Discreet expansion of the LvN in the Fourier space and 2) transferring from Fourier expansion to the Fourier representation. Part one begins by employing discreet Fourier expansions for both the time dependent Hamiltonian and density operators as

$$\hat{H}(t) = \sum_n \hat{H}^{(n)} e^{in\omega t}, \quad \hat{\rho}(t) = \sum_n \hat{\rho}^{(n)}(t) e^{in\omega t}. \tag{S1}$$

Note that the coefficients $\hat{\rho}^{(n)}$ is time-dependent whereas the $\hat{H}^{(n)}$ is not. We then substitute above expansions on the LvN equation, $d\hat{\rho}(t)/dt = -i/\hbar[\hat{H}(t), \hat{\rho}(t)]$, as

$$\sum_n \left( \frac{d\hat{\rho}^{(n)}(t)}{dt} e^{in\omega t} + in\omega \hat{\rho}^{(n)}(t) e^{in\omega t} \right) = -\frac{i}{\hbar} \sum_{k,m} \left[ \hat{H}^{(k)}, \hat{\rho}^{(m)}(t) \right] e^{i(k+m)\omega t} = -\frac{i}{\hbar} \sum_{n,m} \left[ \hat{H}^{(n-m)}, \hat{\rho}^{(m)}(t) \right] e^{in\omega t}. \tag{S2}$$

Next, we introduce the Floquet Number and Floquet Ladder operators as

$$\hat{N}|n\rangle = n|n\rangle, \quad \hat{L}_n|m\rangle = |n+m\rangle. \tag{S3}$$

In the matrix form, $\hat{N}$ can be understood as a matrix with integer numbers on its diagonal and $\hat{L}_n$ is an off-diagonal identity matrix shifted by $n$. Following relations are hold for these two operators

$$[\hat{N}, \hat{L}_n] = n\hat{L}_n, \quad [\hat{L}_n, \hat{L}_m] = 0, \quad \hat{L}_n \hat{L}_m = \hat{L}_m \hat{L}_n = \hat{L}_{n+m}. \tag{S4}$$

Next, we introduce Fourier representations as

$$\hat{H}^f(t) = \sum_n \hat{H}^{(n)} \hat{L}_n e^{in\omega t}, \hat{\rho}^f(t) = \sum_n \hat{\rho}^{(n)}(t) \hat{L}_n e^{in\omega t}, \tag{S5}$$

where we have modified Fourier expansions by adding the ladder operator $\hat{L}_n$. We also introduce the Fourier representation of the LvN equation as

$$\frac{d\hat{\rho}^f(t)}{dt} = -\frac{i}{\hbar}[\hat{H}^f(t), \hat{\rho}^f(t)]. \tag{S6}$$

Afterward, we substitute Fourier representations of $\hat{H}$ and $\hat{\rho}$ [Eqs. (S5)] into above relation as

$$\sum_n \left( \frac{d\hat{\rho}^{(n)}(t)}{dt} \hat{L}_n e^{in\omega t} + in\omega \hat{\rho}^{(n)}(t) \hat{L}_n e^{in\omega t} \right) = -\frac{i}{\hbar} \sum_{k,m} \left[ \hat{H}^{(k)} \hat{L}_k, \hat{\rho}^{(m)}(t) \hat{L}_m \right] e^{i(k+m)\omega t} = -\frac{i}{\hbar} \sum_{n,m} \left[ \hat{H}^{(n-m)} \hat{L}_{n-m}, \hat{\rho}^{(m)}(t) \hat{L}_m \right] e^{in\omega t} = -\frac{i}{\hbar} \sum_{n,m} \left[ \hat{H}^{(n-m)}, \hat{\rho}^{(m)}(t) \right] \hat{L}_n e^{in\omega t}. \tag{S7}$$

where we have used $[\hat{L}_{n-m}, \hat{L}_m] = 0$ and $\hat{L}_{n-m}\hat{L}_m = \hat{L}_n$ in the last line. Since for each $n$, two sides of Eq. (S2) and Eq. (S7) are equivalent then we have proven that the LvN equation in Fourier representations keeps the original form and hence Eq. (S6) is valid.

## DERIVATION OF FOKKER-PLANCK EQUATION

To derive an EOM for the nuclei, we take into account the following weak approximation for the mixed nuclear-electron Floquet density operator $\hat{\rho}_{WF}(\mathbf{R}, \mathbf{P}, t)$ as

$$\hat{\rho}_{WF}(\mathbf{R}, \mathbf{P}, t) = \mathcal{A}(\mathbf{R}, \mathbf{P}, t)\hat{\rho}_{ssF}(\mathbf{R}) + \hat{\mathcal{B}}(\mathbf{R}, \mathbf{P}, t), \tag{S8}$$

where the nuclear phase space density is denoted by $\mathcal{A}(\mathbf{R}, \mathbf{P}, t)$. The Floquet steady-state electronic density operator is denoted by $\hat{\rho}_{ssF}(\mathbf{R})$ and the difference operator is denoted by $\hat{\mathcal{B}}(\mathbf{R}, \mathbf{P}, t)$. Note that $\hat{\mathcal{L}}_{WF}(\hat{\rho}_{ssF}(\mathbf{R})) = 0$ and $\hat{\rho}_{ssF}(\mathbf{R})$ is normalized on the electronic part at all $\mathbf{R}$ such that $Tr_{e,F}(\hat{\rho}_{ssF}(\mathbf{R})) = N$, where $N$ is the Fourier space dimension. For further simplicity, we write a compact form of Eq. (10), the Floquet QCLE, as the following

$$\frac{d}{dt}\hat{\rho}_{WF}(t) = -\hat{\mathcal{L}}_{WF}(\hat{\rho}_{WF}(t)) + \left\{\hat{H}_{WF}, \hat{\rho}_{WF}\right\}_a, \tag{S9}$$

where $\hat{\mathcal{L}}_{WF}(\hat{\rho}_{WF}(t)) \equiv i/\hbar[\hat{H}_{WF}^e, \hat{\rho}_{WF}(t)]$ and $\left\{\hat{A}, \hat{B}\right\}_a \equiv -1/2(\hat{A}\overleftrightarrow{\Lambda}\hat{B} - \hat{B}\overleftrightarrow{\Lambda}\hat{A})$. After substitution of Eq. (S8) in Eq. (S9), and taking the trace over the electronic bath and Fourier space as

$$\frac{\partial}{\partial t}Tr_{e,F}\left(\mathcal{A}(t)\hat{\rho}_{ssF} + \hat{\mathcal{B}}\right) = Tr_{e,F}\left\{\hat{H}_{WF}, \mathcal{A}(t)\hat{\rho}_{ssF}\right\}_a + Tr_{e,F}\left\{\hat{H}_{WF}, \hat{\mathcal{B}}\right\}_a, \tag{S10}$$

we arrive to

$$\frac{\partial}{\partial t}\mathcal{A}(t) = -\sum_\alpha \left(\frac{P_\alpha}{M_\alpha}\right)\frac{\partial \mathcal{A}(t)}{\partial R_\alpha} + \frac{1}{2}\sum_\alpha Tr_{e,F}\left(\frac{\partial \hat{H}_{WF}^e}{\partial R_\alpha}\frac{\partial(\mathcal{A}(t)\hat{\rho}_{ssF})}{\partial P_\alpha} + \frac{\partial(\mathcal{A}(t)\hat{\rho}_{ssF})}{\partial P_\alpha}\frac{\partial \hat{H}_{WF}^e}{\partial R_\alpha}\right) + \\ \frac{1}{2}\sum_\alpha Tr_{e,F}\left(\frac{\partial \hat{H}_{WF}^e}{\partial R_\alpha}\frac{\partial \hat{\mathcal{B}}}{\partial P_\alpha} + \frac{\partial \hat{\mathcal{B}}}{\partial P_\alpha}\frac{\partial \hat{H}_{WF}^e}{\partial R_\alpha}\right). \tag{S11}$$

Note that, $Tr_{e,F}\hat{\mathcal{L}}_{WF}\left(\mathcal{A}(t)\hat{\rho}_{ssF} + \hat{\mathcal{B}}\right) = Tr_{e,F}\hat{\mathcal{L}}_{WF}\left(\hat{\mathcal{B}}\right) = 0$ and $Tr_{e,F}(\partial \mathcal{B}/\partial R^\alpha) = 0$. Since $\hat{\rho}_{ssF}$ does not depends in $P_\alpha$, we can further simplify the above relation as

$$\frac{\partial}{\partial t}\mathcal{A}(t) = -\sum_\alpha \left(\frac{P_\alpha}{M^\alpha}\right)_F \frac{\partial \mathcal{A}(t)}{\partial R_\alpha} + \sum_\alpha Tr_{e,F}\left(\frac{\partial \hat{H}_{WF}^e}{\partial R_\alpha}\hat{\rho}_{ssF}\right)\frac{\partial \mathcal{A}(t)}{\partial P^\alpha} + \sum_\alpha Tr_{e,F}\left(\frac{\partial \hat{H}_{WF}^e}{\partial R_\alpha}\frac{\partial \hat{\mathcal{B}}}{\partial P_\alpha}\right). \tag{S12}$$

In above relation, we have also use the fact that $Tr[AB] = Tr[BA]$. At this point, one needs to express $\hat{\mathcal{B}}$ in terms of $\hat{\mathcal{A}}$. To proceed, we can first have a relation for $\partial \hat{\mathcal{B}}/\partial t$ as:

$$\frac{\partial}{\partial t}\hat{\mathcal{B}} = -\hat{\rho}_{ssF}\frac{\partial}{\partial t}\mathcal{A}(t) + \left\{\hat{H}_{WF}, \mathcal{A}(t)\right\}_a \hat{\rho}_{ssF} + \left\{\hat{H}_{WF}, \hat{\mathcal{B}}\right\}_a - \hat{\mathcal{L}}_{WF}(\hat{\mathcal{B}}) = \\ \left\{\hat{H}_{WF}, \hat{\mathcal{B}}\right\}_a - \hat{\rho}_{ssF}Tr_{e,F}\left\{\hat{H}_{WF}, \hat{\mathcal{B}}\right\}_a - \\ \hat{\rho}_{ssF}Tr_{e,F}\left\{\hat{H}_{WF}, \mathcal{A}(t)\hat{\rho}_{ssF}\right\}_a + \left\{\hat{H}_{WF}, \mathcal{A}(t)\hat{\rho}_{ssF}\right\}_a - \hat{\mathcal{L}}_{WF}(\hat{\mathcal{B}}). \tag{S13}$$

Next, we assume that nuclei move much slower than electrons. With that assumption, only the last three terms of the above relation will survive as

$$\hat{\mathcal{L}}_{WF}(\hat{\mathcal{B}}) = -\hat{\rho}_{ssF}Tr_{e,F}\left\{\hat{H}_{WF}, \mathcal{A}(t)\hat{\rho}_{ssF}\right\}_a + \left\{\hat{H}_{WF}, \mathcal{A}(t)\hat{\rho}_{ssF}\right\}_a = \\ -\hat{\rho}_{ssF}\left(-\sum_\beta \left(\frac{P_\beta}{M_\beta}\right)\frac{\partial \mathcal{A}(t)}{\partial R_\beta} + \sum_\beta Tr_{e,F}\left(\frac{\partial \hat{H}_{WF}^e}{\partial R_\beta}\hat{\rho}_{ssF}\right)\frac{\partial \mathcal{A}(t)}{\partial P_\beta}\right) \\ -\sum_\beta \left(\frac{P_\beta}{M^\beta}\right)\left(\frac{\partial \mathcal{A}(t)}{\partial R_\beta}\hat{\rho}_{ssF} + \frac{\partial \hat{\rho}_{ssF}}{\partial R_\beta}\mathcal{A}(t)\right) \\ + \frac{1}{2}\sum_\beta \left(\frac{\partial \hat{H}_{WF}^e}{\partial R_\beta}\hat{\rho}_{ssF} + \hat{\rho}_{ssF}\frac{\partial \hat{H}_{WF}^e}{\partial R_\beta}\right)\frac{\partial \mathcal{A}(t)}{\partial P_\beta}. \tag{S14}$$

Formal solution for $\hat{\mathcal{B}}$ is given by

$$\hat{\mathcal{B}} = -\sum_\beta \hat{\mathcal{L}}_{WF}^{-1} \frac{\partial \hat{\rho}_{ssF}}{\partial R_\beta} \left( (\frac{P_\beta}{M_\beta}) \mathcal{A}(t) \right)$$

$$-\sum_\beta \hat{\mathcal{L}}_{WF}^{-1} \hat{\rho}_{ssF} Tr_{e,F} \left( \frac{\partial \hat{H}_{WF}^e}{\partial R_\beta} \hat{\rho}_{ssF} \right) \frac{\partial \mathcal{A}(t)}{\partial P_\beta} \quad \text{(S15)}$$

$$+ \frac{1}{2} \sum_\beta \hat{\mathcal{L}}_{WF}^{-1} \left( \frac{\partial \hat{H}_{WF}^e}{\partial R_\beta} \hat{\rho}_{ssF} + \hat{\rho}_{ssF} \frac{\partial \hat{H}_{WF}^e}{\partial R_\beta} \right) \frac{\partial \mathcal{A}(t)}{\partial P_\beta}.$$

Note that the first and third terms of Eq. (S14) cancel each other out. The above relation should be rearranged such that $\partial \hat{\mathcal{B}}_F / \partial P^\alpha$ is given by

$$\frac{\partial \hat{\mathcal{B}}}{\partial P_\alpha} = -\sum_\beta \hat{\mathcal{L}}_{WF}^{-1} \frac{\partial \hat{\rho}_{ssF}}{\partial R_\beta} \frac{\partial}{\partial P_\alpha} \left( (\frac{P_\beta}{M_\beta}) \mathcal{A}(t) \right)$$

$$+ \frac{1}{2} \sum_\beta \hat{\mathcal{L}}_{WF}^{-1} \left( -\hat{\rho}_{ssF} \, 2 Tr_{e,F} \left( \frac{\partial \hat{H}_{WF}^e}{\partial R_\beta} \hat{\rho}_{ssF} \right) + \left( \frac{\partial \hat{H}_{WF}^e}{\partial R_\beta} \hat{\rho}_{ssF} + \hat{\rho}_{ssF} \frac{\partial \hat{H}_{WF}^e}{\partial R_\beta} \right) \right) \frac{\partial}{\partial P_\alpha} \frac{\partial \mathcal{A}(t)}{\partial P_\beta}. \quad \text{(S16)}$$

Substitution of $\partial \hat{\mathcal{B}} / \partial P_\alpha$ in Eq. (S12) give a rise to

$$\frac{\partial}{\partial t} \mathcal{A}(t) = -\sum_\alpha (\frac{P_\alpha}{M_\alpha}) \frac{\partial \mathcal{A}(t)}{\partial R_\alpha} + \sum_\alpha Tr_{e,F} \left( \frac{\partial \hat{H}_{WF}^e}{\partial R_\alpha} \hat{\rho}_{ssF} \right) \frac{\partial \mathcal{A}(t)}{\partial P_\alpha}$$

$$+ \sum_{\alpha,\beta} -Tr_{e,F} \left( \frac{\partial \hat{H}_{WF}^e}{\partial R_\alpha} \hat{\mathcal{L}}_{WF}^{-1} \frac{\partial \hat{\rho}_{ssF}}{\partial R_\beta} \right) \frac{\partial}{\partial P_\alpha} \left( (\frac{P_\beta}{M_\beta}) \mathcal{A}(t) \right) \quad \text{(S17)}$$

$$+ \sum_{\alpha,\beta} \frac{1}{2} Tr_{e,F} \left( \frac{\partial \hat{H}_{WF}^e}{\partial R_\alpha} \hat{\mathcal{L}}_{WF}^{-1} \left( -\hat{\rho}_{ssF} 2 Tr_{e,F} \left( \frac{\partial \hat{H}_{WF}^e}{\partial R_\beta} \hat{\rho}_{ssF} \right) + \frac{\partial \hat{H}_{WF}^e}{\partial R_\beta} \hat{\rho}_{ssF} + \hat{\rho}_{ssF} \frac{\partial \hat{H}_{WF}^e}{\partial R_\beta} \right) \right) \frac{\partial}{\partial P_\alpha} \frac{\partial \mathcal{A}(t)}{\partial P_\beta}.$$

With that, we already derived the Floquet forms of mean force, $F_\alpha$, friction, $\gamma_{\alpha\beta}$, and the correlation function of the random force, $\bar{D}_{\alpha\beta}^S$, by comparing the above relation with the Fokker-Planck equation, Eq. (12), mentioned in the main context [S2]. Comparing the new results with previously derived electronic friction, one can conclude that the Floquet version of electronic friction keeps similar form as the non-Floquet version. However, constituents should be transformed into their Floquet representations.

In the followings, we turn our attention toward further simplifications of the friction tensor. At this point, we employ the identity $\hat{\mathcal{L}}_{WF}^{-1} = \lim_{\eta \to 0^+} \int_0^\infty dt e^{-(\hat{\mathcal{L}}_{WF}+\eta)t}$ where $e^{-\hat{\mathcal{L}}_{WF} t}(A) = e^{-i\hat{H}_{WF}^e t/\hbar}(A) e^{i\hat{H}_{WF}^e t/\hbar}$ ($\eta$ being a positive infinitesimal). With that, the Floquet friction tensor can be given by

$$\gamma_{\alpha\beta} = -\int_0^\infty dt Tr_{e,F} \left( \frac{\partial \hat{H}_{WF}^e}{\partial R_\alpha} e^{-i(\hat{H}_{WF}^e - i\eta)t/\hbar} \frac{\partial \hat{\rho}_{ssF}}{\partial R_\beta} e^{i(\hat{H}_{WF}^e + i\eta)t/\hbar} \right). \quad \text{(S18)}$$

In what follows, $\gamma_{\alpha\beta}$ denotes the Floquet friction tensor without an extra indicator. In fact the above friction relation is expressed in the many-body representation but it holds in the single-particle basis as well. It has been shown previously that if a non-interacting quadratic electronic Hamiltonians of the form $\sum_{ab} \mathcal{H}_{ab}(\mathbf{R}, t) \hat{b}_a^\dagger \hat{b}_b + U(\mathbf{R})$ is considered, then friction relation in the single-particle basis keeps the same format as many-body one [S3]. Similar argument can be repeated to derive the following single-particle alternative of the Floquet friction as

$$\gamma_{\alpha\beta} = -\int_0^\infty dt Tr_{m,F} \left( \frac{\partial \mathcal{H}_F}{\partial R_\alpha} e^{-i(\mathcal{H}_F - i\eta)t/\hbar} \frac{\partial \sigma_{ssF}}{\partial R_\beta} e^{i(\mathcal{H}_F + i\eta)t/\hbar} \right). \quad \text{(S19)}$$

In above relation, $\mathcal{H}_F$ is the Floquet single particle electronic Hamiltonian, and $Tr_{m,F}$ represents the trace over single-particle orbitals and also the trace over Fourier space. Here, we have defined $[\sigma_{ssF}]_{ab} = Tr_e \left( \hat{b}_b^\dagger \hat{b}_a \hat{\rho}_{ssF} \right)$. Note that $U(\mathbf{R})$ does not contribute to the friction.

# FRICTION IN FLOQUET GREEN'S FUNCTION REPRESENTATION

Furthermore, one can recast the Floquet friction tensor into the energy domain as

$$\gamma_{\alpha\beta} = -\int_0^\infty dt \int_0^\infty dt' Tr_{m,F}\left(\frac{\partial \mathcal{H}_F}{\partial R_\alpha} e^{-i(\mathcal{H}_F - i\eta)t/\hbar} \frac{\partial \sigma_{ssF}}{\partial R_\beta} e^{i(\mathcal{H}_F + i\eta)t'/\hbar}\right) \delta(t - t')$$
$$= -\int_{-\infty}^\infty \frac{d\epsilon}{2\pi\hbar} \int_0^\infty dt \int_0^\infty dt' Tr_{m,F}\left(\frac{\partial \mathcal{H}_F}{\partial R_\alpha} e^{-i(\mathcal{H}_F - i\eta)t/\hbar} \frac{\partial \sigma_{ssF}}{\partial R_\beta} e^{i(\mathcal{H}_F + i\eta)t'/\hbar}\right) e^{i\epsilon(t - t')/\hbar} \quad (S20)$$
$$= -\hbar \int_{-\infty}^\infty \frac{d\epsilon}{2\pi} Tr_{m,F}\left(\frac{\partial \mathcal{H}_F}{\partial R_\alpha} \frac{1}{\epsilon + i\eta - \mathcal{H}_F} \frac{\partial \sigma_{ssF}}{\partial R_\beta} \frac{1}{\epsilon - i\eta - \mathcal{H}_F}\right).$$

Hence, we have redefined the $\gamma_{\alpha\beta}$, partially, in terms of the Floquet Retarded and Advanced Green's functions, $G_F^{R/A} = (\epsilon \pm i\eta - \mathcal{H}_F)^{-1}$, as

$$\gamma_{\alpha\beta} = -\hbar \int_{-\infty}^\infty \frac{d\epsilon}{2\pi} Tr_{m,F}\left(\frac{\partial \mathcal{H}_F}{\partial R_\alpha} G_F^R \frac{\partial \sigma_{ssF}}{\partial R_\beta} G_F^A\right). \quad (S21)$$

For a practical calculation of electronic friction tensors, one needs to express the derivative $\partial \sigma_{ssF}/\partial R_\beta$ in terms of the Floquet lesser Green's function denoted by $G_F^<$. The $\sigma_{ssF}$ relates to the $G_F^<$ by

$$\sigma_{ssF} = \int \frac{d\epsilon'}{2\pi i} G_F^<(\epsilon') = \int \frac{d\epsilon'}{2\pi i} G_F^R(\epsilon') \Sigma_F^<(\epsilon') G_F^A(\epsilon'), \quad (S22)$$

where $\Sigma_F^<$ is the total lead's Floquet lesser self-energy[S4]. Here, we have adopted a dot-lead (system-bath) separation. Furthermore, we have assumed that $\Sigma_F^<$ neither depends on the energy $\epsilon'$ (so-called wide-band approximation) nor on the position $\mathbf{R}$ (so-called Condon approximation). Note that the wide-band approximation allows us to express the lesser green's function in the Floquet representation as $G_F^<(\epsilon') = G_F^R(\epsilon')\Sigma_F^< G_F^A(\epsilon')$. With Condon approximation, one can easily derive the following identity

$$\frac{\partial G_F^<}{\partial R_\beta} = G_F^R \frac{\partial \mathcal{H}_F}{\partial R_\beta} G_F^< + G_F^< \frac{\partial \mathcal{H}_F}{\partial R_\beta} G_F^A. \quad (S23)$$

It is important to note that we can not directly substitute $\partial \sigma_{ssF}/\partial R_\beta$ of Eq. (S22) into Eq. (S21) due to extra integration over $\epsilon'$. To proceed further, we can replace the $Tr_{m,F}(...)$ with $\sum_n \langle n|...|n\rangle$ in the last line of Eq. (S20) and use the eigenbasis of the Floquet electronic Hamiltonian, $\mathcal{H}_F|n\rangle = \epsilon_{nF}|n\rangle$, as

$$\gamma_{\alpha\beta} = -\hbar \sum_n \int_{-\infty}^\infty \frac{d\epsilon}{2\pi} \langle n|\frac{\partial \mathcal{H}_F}{\partial R_\alpha} \frac{1}{\epsilon + i\eta - \mathcal{H}_F} \frac{\partial \sigma_{ssF}}{\partial R_\beta}|n\rangle \frac{1}{\epsilon - i\eta - \epsilon_{nF}}. \quad (S24)$$

Next, we will use the Floquet identity operator $\sum_m |m\rangle\langle m|$ as

$$\gamma_{\alpha\beta} = -\hbar \sum_{n,m} \int_{-\infty}^\infty \frac{d\epsilon}{2\pi} \langle n|\frac{\partial \mathcal{H}_F}{\partial R_\alpha}|m\rangle \frac{1}{\epsilon + i\eta - \epsilon_{mF}} \langle m|\frac{\partial \sigma_{ssF}}{\partial R_\beta}|n\rangle \frac{1}{\epsilon - i\eta - \epsilon_{nF}}. \quad (S25)$$

Taking the singularity point at $\epsilon = i\eta + \epsilon_{nF}$ and using the residue theorem for contour integration leads to

$$\gamma_{\alpha\beta} = -i\hbar \sum_{n,m} \langle n|\frac{\partial \mathcal{H}_F}{\partial R_\alpha}|m\rangle \frac{1}{\epsilon_{nF} - \epsilon_{mF} + i2\eta} \langle m|\frac{\partial \sigma_{ssF}}{\partial R_\beta}|n\rangle. \quad (S26)$$

At this point, we will evaluate the last term of above expression as: $\langle m|\frac{\partial \sigma_{ssF}}{\partial R_\beta}|n\rangle$. According to Eqs. (S22) and (S23), this term has two parts as

$$\langle m|\frac{\partial \sigma_{ssF}}{\partial R_\beta}|n\rangle = \int \frac{d\epsilon'}{2\pi i} \langle m|G_F^R(\epsilon')\frac{\partial \mathcal{H}_F}{\partial R_\beta} G_F^R(\epsilon')\Sigma_F^< G_F^A(\epsilon')|n\rangle + \langle m|G_F^R(\epsilon')\Sigma_F^< G_F^A(\epsilon')\frac{\partial \mathcal{H}_F}{\partial R_\beta} G_F^A(\epsilon')|n\rangle. \quad (S27)$$

The integration over $\epsilon'$ can be accomplished by using the eigenbasis of the Floquet electronic Hamiltonian and employing the identity operator $\sum_{m'} |m'\rangle\langle m'|$. The first part is given by

$$\sum_{m'} \int \frac{d\epsilon'}{2\pi i} \frac{1}{\epsilon' + i\eta - \epsilon_{mF}} \langle m|\frac{\partial \mathcal{H}_F}{\partial R_\beta}|m'\rangle \frac{1}{\epsilon' + i\eta - \epsilon_{m'F}} \langle m'|\Sigma_F^<|n\rangle \frac{1}{\epsilon' - i\eta - \epsilon_{nF}}$$
$$= \sum_{m'} \frac{1}{\epsilon_{nF} - \epsilon_{mF} + i2\eta} \langle m|\frac{\partial \mathcal{H}_F}{\partial R_\beta}|m'\rangle \frac{1}{\epsilon_{nF} - \epsilon_{m'F} + i2\eta} \langle m'|\Sigma_F^<|n\rangle. \quad (S28)$$

Similarly (by taking the singularity point at $\epsilon' = -i\eta + \epsilon_{mF}$), the second part reduces to

$$\sum_{m'} \langle m|\Sigma_F^<|m'\rangle \frac{1}{\epsilon_{mF} - \epsilon_{m'F} - i2\eta} \langle m'|\frac{\partial \mathcal{H}_F}{\partial R_\beta}|n\rangle \frac{1}{\epsilon_{mF} - \epsilon_{nF} - i2\eta}. \quad (S29)$$

A relation for $\gamma_{\alpha\beta}$ can be derived by substitution of these two parts in the Eq. (S25) as

$$\gamma_{\alpha\beta} = -i\hbar \sum_{n,m,m'} \langle n|\frac{\partial \mathcal{H}_F}{\partial R_\alpha}|m\rangle \frac{1}{\epsilon_{nF} - \epsilon_{mF} + i2\eta}$$
$$\left( \frac{1}{\epsilon_{nF} - \epsilon_{mF} + i2\eta} \langle m|\frac{\partial \mathcal{H}_F}{\partial R_\beta}|m'\rangle \frac{1}{\epsilon_{nF} - \epsilon_{m'F} + i2\eta} \langle m'|\Sigma_F^<|n\rangle \right.$$
$$\left. + \langle m|\Sigma_F^<|m'\rangle \frac{1}{\epsilon_{mF} - \epsilon_{m'F} - i2\eta} \langle m'|\frac{\partial \mathcal{H}_F}{\partial R_\beta}|n\rangle \frac{1}{\epsilon_{mF} - \epsilon_{nF} - i2\eta} \right). \quad (S30)$$

Taking similar procedures (replacing $Tr_{m,F}(...)$ by $\sum_n \langle n|...|n\rangle$, using the eigenbasis of the Floquet electronic Hamiltonian and employing the identity operators) one can conclude the following general single integration formula

$$\gamma_{\alpha\beta} = \hbar \int_{-\infty}^{\infty} \frac{d\epsilon}{2\pi} \, Tr_{m,F} \left( \frac{\partial \mathcal{H}_F}{\partial R_\alpha} \frac{\partial G_F^R}{\partial \epsilon} \frac{\partial \mathcal{H}_F}{\partial R_\beta} G_F^< - \frac{\partial \mathcal{H}_F}{\partial R_\alpha} G_F^< \frac{\partial \mathcal{H}_F}{\partial R_\beta} \frac{\partial G_F^A}{\partial \epsilon} \right), \quad (S31)$$

delivers a similar outputs as Eq. (S30). Note that, we have used the identity $\partial G_F^{R,A}/\partial \epsilon = -G_F^{R,A} G_F^{R,A}$ [S2]. This relation is a practical formula for evaluation of $\gamma_{\alpha\beta}$. The second term in Eq. (S30) is also the hermitian conjugate of the first part. Since the trace in any basis set is the same, Eq. (S31) represents a general form for Floquet electronic friction.


\fi

\end{document}